\documentclass[12pt,onecolumn,romanappendices]{IEEEtran}
\usepackage[caption=false]{subfig}
\usepackage{stfloats}
\IEEEoverridecommandlockouts
\newcommand{\Fb}{{\mathbb F}}

\newcommand{\nonl}{\renewcommand{\nl}{\let\nl\oldnl}}% Remove line number for one line
\usepackage{etoolbox}
\AtEndEnvironment{example}{\null\hfill$\square$}%
\AtEndEnvironment{remark}{\null\hfill$\square$}%

\usepackage{enumerate}
\usepackage{enumitem}
\usepackage{multirow,array}
\usepackage{cite}
\usepackage{graphicx}
\usepackage{psfrag}
\usepackage{array}
\usepackage{algorithm2e}
\usepackage{amssymb}
\usepackage{amsfonts}
\usepackage{float}
\usepackage{textcomp}
\usepackage{xcolor}
\usepackage{tabu}
\usepackage{array}
\usepackage{tabularx}
\usepackage{subfiles}

\usepackage{amssymb}% http://ctan.org/pkg/amssymb
\usepackage{pifont}% http://ctan.org/pkg/pifont

\newcolumntype{P}[1]{>{\centering\arraybackslash}p{#1}}
\newcolumntype{M}[1]{>{\centering\arraybackslash}m{#1}}
\usepackage{amsmath}
\usepackage{tikz}
\usepackage{comment}
\usepackage{amsthm}
\theoremstyle{remark}

\newtheorem*{remark*}{Remark}
\usepackage{float}
\newtheorem{conjecture*}{Conjecture}
\newtheorem{claim*}{Claim}

\newtheorem{corollary}{Corollary}
\newtheorem{claim}{Claim}

\newtheorem{theorem}{Theorem}
\newtheorem{example}{Example}

\newcommand{\pkarxiv}[1]{}
\newcommand{\capac}{\mathsf{C}}

\title{Private Information Retrieval for Graph-based Replication  with Minimal Subpacketization}

\begin{document}
%%%%%%%%%%%%%%%%%%%%%%%%%%%%%%%%%%%%%%%%%%%%%%%%%%%%%%%%%%%%
% \author{
%   \IEEEauthorblockN{Authors}

% \vspace{-0.2cm}
% %%%%%%%%%%%%%%%%%%%
% }

\author{Vayur Shanbhag, Prasad Krishnan % <- this % stops a space

\thanks{Vayur and Dr. Krishnan are with the Signal Processing and Communications Research Center, International Institute of Information Technology, Hyderabad, 500032, India (email: $\{$vayur.s@research., prasad.krishnan@$\}$iiit.ac.in).
% Acknowledgment:  Dr. Krishnan acknowledges support from ANRF-SERB project CRG/2023/008696. 
}
}

\maketitle

\allowdisplaybreaks % Command for 

%%%%%%%%%%%%%%%%%%%%%%%%%%%%%%%%%%%%%%%%%%%%%%%%%%%%%%%%%%%%
\begin{abstract}
% THIS PAPER IS ELIGIBLE FOR THE STUDENT PAPER AWARD. 
We design new minimal-subpacketization schemes for information-theoretic private information retrieval on graph-based replicated databases. In graph-based replication, the system consists of $K$ files replicated across $N$ servers according to a graph with $N$ vertices and $K$ edges. The client wants to retrieve one desired file, while keeping the index of the desired file private from each server via a query-response protocol. We seek PIR protocols that have (a) high rate, which is the ratio of the file-size to the total download cost, and (b) low subpacketization, which acts as a constraint on the size of the files for executing the protocol. We report two new schemes which have unit-subpacketization (which is minimal): (i) for a special class of graphs known as star graphs, and (ii) for general graphs. Our star-graph scheme has a better rate than previously known schemes with low subpacketization for general star graphs. Our scheme for general graphs uses a decomposition of the graph via independent sets. This scheme achieves a rate lower than prior schemes for the complete graph, however it can achieve higher rates than known for some specific graph classes. An extension of our scheme to the case of multigraphs achieves a higher rate than previous schemes for the complete multi-graph.
\end{abstract}
\section{Introduction}
\label{sec:intro}

Consider the scenario where a client wants to retrieve a desired file from a database containing $K$ files  stored across $N$ \lq honest-but-curious' servers. A private information retrieval (PIR) scheme \cite{chor1998_PIR} is a protocol between the client and the servers, in which the client sends a query to each server and receives a response from each. We require that: (a) from the answers the client receives, it is able to decode the desired file (correctness), and (b) the identity of the desired file is not revealed to any server (privacy). Formally, we define the \textit{rate} of a scheme as the ratio of the file size to the download cost. Denoted by $\capac$, the supremum of this ratio over all file sizes, and over all schemes is the capacity of PIR (for a particular storage configuration). During retrieval, each file is divided into $L$ parts, and this is known as the \lq subpacketization' of the scheme. The size of the files must be at least $L$ bits, to ensure that the scheme can be employed. It is desirable to have PIR schemes whose rate is close to the capacity, and have small subpacketization. For the setting of replicated storage, the capacity of PIR was characterized in \cite{sun2017capacity}.  Following this, a number of works studied and characterized capacity of problem for other scenarios, such as for coded databases \cite{banawan2018capacity,friejHollantiSIAM,PIR_linearcodes_arbit,tPIRIEEE} and for server-collusion \cite{sun2017capacityb}.

PIR schemes in the literature can be broadly classified into two classes: \textit{fixed-download} schemes and \textit{variable-download} schemes. In fixed-download schemes, the quantum of download is the same across instances of the protocol, whereas in variable-download schemes, the quantum of download is determined by some private randomness generated by the client, and varies across instances of the protocol. Variable-download schemes tend to have lower subpacketization than fixed-download schemes. 
%The total download in a variable-download scheme is calculated as the expectation of the total download, with respect to the distribution of the client's private randomness. 
A variable-download capacity-achieving scheme with optimal subpacketization was given in~\cite{tian2019capacity}.

Under the file-replication setup, of special interest is the case where each server stores not all but only a subset of files, as this lowers the storage-cost per server. 
% Several works (eg~) exist for the case where all servers do not hold the entire database. %These include both coded and uncoded versions. 
Among the works which studied PIR for such storage-constrained databases (for instance, \cite{Tian2020_storageCost, ravivetal_2018_TIT,attia_kumar_tandon_storageconstrained_TIT}), the work \cite{ravivetal_2018_TIT} introduced the notion of \textit{graph-based replication}. In this model, we assume that the files are stored uncoded, and each file is stored on exactly two servers. Such a storage configuration can be modeled as a graph, where each vertex corresponds to a server, and each edge is a file stored on the servers incident to it. The work \cite{ravivetal_2018_TIT} presented achievability and converse results on PIR for (hyper)-graph based replication, for both colluding and non-colluding setups. Following the model introduced in \cite{ravivetal_2018_TIT}, a variety of results have been obtained \cite{Sadehetal_TIT_2023_BoundsonPIRgraphs,ge2025privateinformationretrievalgraphs,Kong_TIT_2025newcapacityboundspir,Yuhangetal_Capof4StarGraph_ISIT_2023,meel2025privateinformationretrievalmultigraphbased}. The focus of the prior works has been to characterize the capacity of PIR for general and specific classes of graphs, via achievability schemes and information-theoretic upper bounds.   

 %(although~\cite{ge2025privateinformationretrievalgraphs} comes very close to a resolution). 

The complete graph on $N$ vertices, denoted by $K_N$ is of special interest. This is because C($K_N$) is a lower bound on the capacity of any graph with $N$ vertices \cite{ge2025privateinformationretrievalgraphs}.
Another well-studied special category of graphs is the star graph~\cite{Sadehetal_TIT_2023_BoundsonPIRgraphs,Yuhangetal_Capof4StarGraph_ISIT_2023,Kong_TIT_2025newcapacityboundspir}, which has a single central (`hub') server and $N-1$ `spoke' servers which connect to the same. The total number of files on this is $N-1$. Star graphs are of interest since any graph is a union of star graphs. See Table~\ref{table:pir_graphs_previousWork} for a summary of prior work, listing various (and perhaps non-exhaustive) known results for general graphs and also for special graph structures, along with the subpacketization required for the achievability results (which give lower bounds on the capacity). While the capacity of general graphs is not tightly characterized, it can be seen from Table \ref{table:pir_graphs_previousWork} that the capacity is known for some cases, and for few others, a tight characterization is also known. The work \cite{Kong_TIT_2025newcapacityboundspir} introduced the $r$-multigraph extension of a graph, in which each edge is replaced by $r$ parallel edges, i.e., where pairs of servers share $r$ files for $r\geq 1$. 

% Consequently, the number of files increases by a multiplicative factor of $r$.  for a graph $G$, we denote its $r$-multigraph extension by $G^{(r)}$. 

In this work, we present novel unit-subpacketization ($L=1$, which is minimal) variable-download PIR schemes for the star-graph and for general graphs. While having minimal subpacketization, our schemes have also have better rates over the prior schemes, in some cases. The specific contributions and organization of this work are as follows. 
%%%%% 
\begin{itemize}[leftmargin=*]
\item After formally reviewing the system model in Section \ref{sec:systemmodel}, we provide our new PIR protocol for star graphs in Section \ref{sec:starscheme1}. Our scheme on star graphs achieves $R \geq  \frac{1}{2\sqrt{N} -2 + \frac{1}{\sqrt{N}+ 1}}$ (if $N$ is not a perfect square, we replace it with the smallest perfect square that is larger). The rate achieved is better than the scheme in \cite{Sadehetal_TIT_2023_BoundsonPIRgraphs} for any general star graph. It is slightly worse than the capacity of the $4$-star graph with $N=5$ shown in \cite{Yuhangetal_Capof4StarGraph_ISIT_2023}. Our subpacketization $L=1$ is better than both schemes. 
\item On general graphs, the scheme we present in Section \ref{sec:generalgraph1} achieves $R \geq \max\left(\frac{2}{2N - \alpha(G)}, \frac{1}{N-1}\right)$, where $\alpha(G)$ is the size of the largest independent set of the graph. This rate is smaller than \cite{Sadehetal_TIT_2023_BoundsonPIRgraphs, Kong_TIT_2025newcapacityboundspir, ge2025privateinformationretrievalgraphs} on complete graphs. However, for complete bipartite graphs, our scheme achieves $R \geq 4/3N$ which is the best in the literature. We extend our scheme to $r$-multigraphs (Subsection \ref{subsec:multigraph}), thus obtaining rate $R = \frac{1}{N - 2^{1-r}}$ for the complete multigraph extension of $K_N$, which is the best in the literature.  We conclude our findings with a summary in Section~\ref{sec:conclusion}.
\end{itemize}

% The rest of the work is organized as follows. The remainder of this section is devoted to relevant notations. We present the system model and relevant literature in Section~\ref{sec:systemmodel}. In Section~\ref{sec:starscheme1} we describe a scheme for star graphs.
% In Section~\ref{sec:generalgraph1} we describe a scheme for general graphs and its extension to multigraphs. We conclude our findings in Section~\ref{sec:conclusion}.

\textit{Notation:} Let $[N]$ denote the set $\{1,\hdots,N\}$, for any positive integer $N$. The collection of quantities $A_n:n\in[N]$ is denoted as $A_{[N]}$. A (simple, undirected) graph $G$ is defined by its vertex set $V$ and edge set $E$, where $E\subseteq V\times V$. 
% A graph $H$ is a subgraph of graph $G$, denoted by $H\subseteq G$, if the sets of vertices and edges of $H$ are subsets of those of $G$, respectively. 
The subgraph of $G$ induced by a subset $S$ of its vertices is denoted by $G[S]$. A subset $S\subset V$ of the vertices of $G$ is said to be an independent set of $G$, if no pair of vertices of $S$ have an edge between. The largest size of an independent set of $G$ is called the independence number of $G$, and is denoted by $\alpha(G)$. 
%The degree of the graph, denoted by $\Delta(G)$, is the maximum degree of any vertex in $G$. 
The complete graph on $N$ vertices is denoted by $K_N$. The complete bipartite graph with $N_1$ vertices on one-side and $N_2$ vertices on the other is denoted by $K_{N_1,N_2}$. The notation $\emptyset$ denotes the empty set. For sets, $A, B$ we define $A \setminus B$ as the set of elements in $A$ but not in $B$. The binary field is denoted as $\Fb_2$. Let $\overline{1}$ denote the all-ones vector and $\overline{0}$ denote the all-zeros vector over any finite field (their lengths will be clear from the context). %\pk{Check if needed: In some places, we use the abbreviation `s.t.' to denote the phrase `such that'.} 
The $(i,j)^{\text{th}}$ entry of a matrix $M$ is written as $M(i,j)$.

\begin{table*}[t]
\centering
\captionsetup{justification=centering}
\caption{\small Summary of known capacity bounds in previous works. Results marked with $^\dagger$ are state-of-the-art achievable rates so far. }
\label{table:pir_graphs_previousWork}

\setlength{\tabcolsep}{4pt}
\renewcommand{\arraystretch}{1.4}

\scriptsize
\begin{tabularx}{\textwidth}{|X|X|X|p{3cm}|}
\hline
\textbf{Graphs and Scheme} 
& \textbf{Achievable Rate (lower bound)} 
& \textbf{Subpacketization} 
& \textbf{Known Upper Bound} \\
\hline

$\dagger$ Star graph: fixed-download \cite[Theorem 18]{Sadehetal_TIT_2023_BoundsonPIRgraphs}
& $\approx \tfrac{1}{2\sqrt{N-1}+1}$
& $\sqrt{N-1}+1$
& $\tfrac{1}{\sqrt{2N}-1}$ \\
\hline

$^\dagger$ 4-Star graph: fixed-download \cite[Theorem 1]{Yuhangetal_Capof4StarGraph_ISIT_2023}
& $\tfrac{5}{12}$
& $5$
& $\tfrac{5}{12}$ \\
\hline

$\dagger$ Multi-star graph: fixed-download ($r>1$)  \cite[Theorem IV.1]{Kong_TIT_2025newcapacityboundspir}
& $\tfrac{2}{N}\!\left(2-\tfrac{1}{2^{r-1}}\right)^{-1}$
& $2^{r}$
& $(2-2^{1-r})^{-1}$ \\
\hline

Any graph: fixed-download \cite[Section III]{ravivetal_2018_TIT}
& $\tfrac{1}{N}$
& $1$
& $\tfrac{2}{N}$ \\
\hline

Complete graph: fixed-download \cite[Theorem 22]{Sadehetal_TIT_2023_BoundsonPIRgraphs}
& $\tfrac{2^{N-1}}{2^{N-1}-1}\tfrac{1}{N}$
& $2^{N-1}$
& $\tfrac{2}{N+1}$ \\
\hline

Complete graph: fixed-download \cite[Construction 2]{Kong_TIT_2025newcapacityboundspir}
& $\tfrac{6}{5-2^{3-N}}\tfrac{1}{N}$
& $3\cdot 2^{N-2}$
& $\tfrac{2}{N+1}$ \\
\hline

$^\dagger$Complete graph (applies to any graph): fixed-download \cite[Theorem 1.12]{ge2025privateinformationretrievalgraphs}
& $\approx (1-o(1))\tfrac{4}{3N}$
& $(N!)^{O(1)}$ (reducible to $1$,via variable-download)% \vspace{-0.4cm}
& $\tfrac{1}{\sum_{i=2}^N\tfrac{1}{i!}}\tfrac{1}{N}$ \\ 
\hline

Complete multigraph: fixed-download \cite[Theorem IV.1]{Kong_TIT_2025newcapacityboundspir}
& $\tfrac{6}{5-2^{3-N}}\tfrac{1}{N}
\!\left(2-\tfrac{1}{2^{r-1}}\right)^{-1}$
& $3\cdot 2^{N+r-3}$
& $\tfrac{1}{N-(N-1)r^{-r}}$ \\
\hline

$^\dagger$ Complete multigraph: variable-download \cite[Sec. 4.2.2]{ge2025privateinformationretrievalgraphs}
& $\tfrac{1}{N-\tfrac{N}{2^{r(N-1)}}}$
& $1$
& (given above) \\
\hline

$\dagger$ Complete bipartite graph ($K_{N_1,N_2}$, $N=N_1+N_2$) 
\cite[Theorem III.4]{Kong_TIT_2025newcapacityboundspir},
\cite[Theorem 1.11]{ge2025privateinformationretrievalgraphs}
& $ \text{Max} \begin{cases}
    \tfrac{6}{5-2^{3-(N_1+N_2)}}\tfrac{1}{N_1+N_2} \\
    \tfrac{1}{2N_2\sqrt{N_1}+N_2}
\end{cases}
$
& $3\cdot 2^{N-2}$ or $\sqrt{N_1}+1$
& $\displaystyle
\frac{1}{N \left(\sum_{i=1}^{N/2} \tfrac{1}{i!2^i}\right)},$ \\
& & & $ N_1 = N_2 = N/2 $ \\
\hline

$^\dagger$ Cycle graph: variable-download \cite[Theorem 1]{2020_IWCIT_OptimalMsgSizePIRnonrepl_keramaatietal}
& $\tfrac{2}{N+1}$
& $1$
& $\tfrac{2}{N+1}$ \\
\hline

\end{tabularx}
\end{table*}
\section{System Model and Relevant Literature}
\label{sec:systemmodel}
\subsection{System model}
We consider a distributed storage system on $N$ storage nodes, which store messages according to a graph-based replication scheme, following literature \cite{ravivetal_2018_TIT,Sadehetal_TIT_2023_BoundsonPIRgraphs}. The storage graph is a simple graph $G$, whose vertex set $[N]$ represents the set of $N$ servers, and the edge set $E$ denotes the set of indices of the messages stored in the $N$ servers. That is, an edge $\{i,j\}\in E$ (where $i,j\in[N]$) denotes the message (or file) $W_{i,j}$ (also equivalently represented as $W_{j,i}$) replicated in servers $i$ and $j$. Thus, $W_{i,j}:\{i,j\}\in E$ denotes the set of all messages in the storage system. We assume that the messages $W_{i,j}:\{i,j\}\in E$ are independent and identically distributed, with $H(W_{i,j})=L$ bits, where $L$ denotes the subpacketization (we assume each of the $L$ parts of each file constitutes a single bit). 

The client seeks to retrieve one file $W_\theta$ from these servers privately (i.e., without revealing $\theta$ to any server), where we assume $\theta$ is chosen uniformly at random from all file-indices. A private information retrieval (PIR) protocol executed by the client to retrieve the file involves queries sent to the servers by the client, and the corresponding responses from the servers. Let $Q_n$ denote the query sent to server $n\in [N]$. As in prior work, we model $Q_n$ as a function of desired file index $\theta$ and some private randomness at the client, taking values in some set ${\cal Q}_n$. Further, as the client has no access to $W_{i,j}:\{i,j\}\in E$, thus we have $I(\{W_{i,j}:\{i,j\}\in E\};Q_{[N]})=0$.  The response sent back to the client by server $n$, denoted by $A_n$, is a function of the files stored at server $n$ and the query received $Q_n$. A valid PIR protocol must satisfy two conditions: (a) the \textit{correctness condition}, i.e., $H(W_\theta|Q_{[N]},A_{[N]})=0$, and (b) the \textit{privacy condition} $I(\theta;Q_n,A_n)=0$, for each $n\in[N]$.

The rate of such a PIR protocol is defined as $R\triangleq \frac{L}{\sum_{n\in[N]}H(A_n|Q_n)}.$ For the given graph $G$ which defines the storage scheme, the maximum rate of any PIR protocol for this graph is defined as its capacity, i.e., $\capac(G)\triangleq \sup R$, where the supremum is across all possible PIR protocols over all possible values of $L$. The subpacketization $L$ can be thought of as a measure of complexity of the protocol. Thus, we would like PIR schemes with large rates and small $L$. Clearly, the minimal $L$ possible is $L=1$. 
%%%%
\subsection{Summary of relevant literature} For complete graphs, and by extension for general graphs, 
it was proved that $\capac(K_N)\geq \frac{2^{N-1}}{2^{N-1}-1}\cdot\frac{1}{N}$ in~\cite{Sadehetal_TIT_2023_BoundsonPIRgraphs}. This was then improved to $\frac{6}{5-2^{3-N}}\cdot\frac{1}{N}$ in \cite{Kong_TIT_2025newcapacityboundspir}, which was improved to $(\frac{4}{3} - o(1))\frac{1}{N}$ in \cite{ge2025privateinformationretrievalgraphs}. The tightest upper bound $\capac(K_N) \leq \tfrac{1}{\sum_{i=2}^N\tfrac{1}{i!}}\tfrac{1}{N}$ is given in~\cite{ge2025privateinformationretrievalgraphs}.

%\pk{have to check where to put this}

For star graphs, the work \cite{Sadehetal_TIT_2023_BoundsonPIRgraphs} presents a scheme which achieves a rate of $R \geq \frac{1}{2\sqrt{N-1} + 1}$ with $L = \sqrt{N-1} + 1$, for any star graph (if $N-1$ is not a square then it achieves rate $\Theta(\frac{1}{\sqrt{N}})$). This is also shown to be orderwise optimal in \cite{Sadehetal_TIT_2023_BoundsonPIRgraphs}. The capacity upper bound due to \cite{Sadehetal_TIT_2023_BoundsonPIRgraphs} is $\capac \leq \frac{1}{\sqrt{2N} -1}$. Another scheme for star graphs is given in \cite[Section III]{Yuhangetal_Capof4StarGraph_ISIT_2023}. This scheme achieves capacity $C = 5/12$ for $K = 4$. The exact rate achieved for higher $K$ is not clearly mentioned in \cite{Yuhangetal_Capof4StarGraph_ISIT_2023}, but the scheme uses subpacketization $L \approx \frac{1}{\sqrt{K}} \binom{K}{\sqrt{K}}$. 
 % and rate $R \approx \frac{1}{2\sqrt{K} - 3}$ with $L \approx \frac{1}{\sqrt{K}} \binom{K}{\sqrt{K}}$ asymptotically for large $K$ (the actual rate is not precisely available in \cite{Yuhangetal_Capof4StarGraph_ISIT_2023}).
 
On complete bipartite graphs($K_{N_1,N_2}$), the lower bound $R = \max\!\left\{
\tfrac{6}{5-2^{3-(N_1+N_2)}}\tfrac{1}{N_1+N_2},
\tfrac{1}{2N_1\sqrt{N_2}+N_1}
\right\}$ is derived using the best case between the scheme for complete graphs, and scheme for star graphs repeated multiple times~[Theorem III.4]\cite{Kong_TIT_2025newcapacityboundspir}
and the upper bound \cite[Theorem 1.11]{ge2025privateinformationretrievalgraphs} is $1.5415/N$. 

The previous best lower bound on $r$-multigraph extension of the complete graph, denoted by $K_N^{(r)}$, is given in \cite[Theorem 1.15]{ge2025privateinformationretrievalgraphs} by $R = \tfrac{1}{N-\tfrac{N}{2^{r(N-1)}}}$ and the tightest upper bound in \cite{Kong_TIT_2025newcapacityboundspir} by $\capac(K_N^{(r)}) \leq \tfrac{1}{N-(N-1)r^{-r}}$. The relevant subpacketizations for all these schemes is shown in Table \ref{table:pir_graphs_previousWork}. 
Any fixed-download scheme satisfying an `independence property' can be converted \cite[Theorem 1.14]{ge2025privateinformationretrievalgraphs} into a variable-download, unit-subpacketization scheme with the same rate. This is applicable to schemes on complete graphs (and thus apply to general graphs), but not so for specialized schemes on general star graphs. However, this requires a large query complexity when the subpacketization of the original scheme is large. 

In the subsequent sections, we provide our unit-subpacketization variable-download PIR schemes for the star-graph and for general graphs. Our schemes have advantages over the prior schemes, in terms of both rate and subpacketization in some cases. 
%It is unclear whether the scheme can be converted into one with a lower subpacketization using techniques discussed in \cite{ge2025privateinformationretrievalgraphs}.

\section{A New High-Rate PIR scheme for Star Graphs}
\label{sec:starscheme1}
We consider the star graph on $N$ vertices.  Denote the number of files by $K$. We have $K = N -1$. Let
$\mathcal{S} = \{S_1, S_2, \dots, S_N\} $
denote the set of servers where $S_N$ is the central (\textit{hub}) server and $S_{1:N-1}$ are the $N-1$ \textit{spoke} servers.  Let
$\mathcal{W} = \{W_1, \dots , W_{N-1}\}$
denote the set of all files, and assume $W_i$ is the file stored at servers $S_i$ and $S_N$.
The following theorem captures the rate of a novel variable-download scheme that we describe in this section. 

\begin{theorem}\label{thm:starvar}
    There exists a variable-download PIR scheme for star graphs on $N$ vertices (i.e, with $N-1$ files) that employs subpacketization $L = 1$ and achieves a rate of $R=\Theta\left(\frac{1}{\sqrt{N}}\right)$. In particular, letting $N'$ be the smallest integer such that $N'$ is a perfect square and $N'\geq N$, the scheme achieves a rate $R \geq  \frac{1}{2\sqrt{N'} -2 + \frac{1}{\sqrt{N'}+ 1}}$. 
    
    % There exists a variable-download PIR scheme for star graphs on $K+1$ vertices that achieves a rate of $R = \frac{1}{2\sqrt{K+1} -2 + \frac{1}{\sqrt{K+1} + 1}}$ in the best case. The scheme has subpacketization $L = 1$. 
\end{theorem}

The rest of this section is devoted to the proof of Theorem \ref{thm:starvar}, via the description of the scheme, verifying its correctness and privacy, and finally calculating the rate. The key idea of the scheme is as follows. 
\begin{itemize}[leftmargin=*]
    \item For some non-negative integer $u$ such that $(u+1)|K$, the client downloads $u$ files from a random subset of $u$ spoke servers. 
    \item If the desired file $W_\theta$ is among the files downloaded, then nothing remains to be done. Else, the downloaded files serve as side-information. The client subsequently downloads appropriate linear combinations of all files from the hub server, utilizing available side-information to decode the file while maintaining privacy. The rate achieved is $(u+1)/(u^2+K)$. Optimizing $u$ gives us Theorem \ref{thm:starvar}.
\end{itemize}
%%%
 We now remark on the comparison with the earlier results from \cite{Sadehetal_TIT_2023_BoundsonPIRgraphs, Yuhangetal_Capof4StarGraph_ISIT_2023} (See Table \ref{table:pir_graphs_previousWork}). 
\begin{remark*}[Comparison with prior work]
    The scheme in \cite{Sadehetal_TIT_2023_BoundsonPIRgraphs} achieves $R = \frac{1}{2\sqrt{N'-1} -1  + \frac{1}{\sqrt{N'-1}+ 1}}$ and uses subpacketization $L=\Theta(\sqrt{N})$. The scheme in \cite{Yuhangetal_Capof4StarGraph_ISIT_2023} potentially achieves a rate  which is better than \cite{Sadehetal_TIT_2023_BoundsonPIRgraphs} and Theorem~\ref{thm:starvar} (the precise expression is missing in \cite{Yuhangetal_Capof4StarGraph_ISIT_2023}). However, it requires $L = \binom{K-1}{t-1} + \binom{K-2}{t-2}$, where $t$ is a parameter close to $\sqrt{K}$ (ie. $L \approx \frac{1}{\sqrt{K}} \binom{K}{\sqrt{K}}$) (note, $K=N-1$). Thus, for general star graphs, Theorem \ref{thm:starvar} achieves better subpacketization than prior work, and a rate larger than \cite{Sadehetal_TIT_2023_BoundsonPIRgraphs}.  Note that the upper bound on the capacity shown in \cite{Sadehetal_TIT_2023_BoundsonPIRgraphs} is $1/(\sqrt{2(N-1)}-1)$. For the star graph with $K=4$, the scheme from \cite{Yuhangetal_Capof4StarGraph_ISIT_2023} achieves rate $5/12$ with subpacketization $5$, whereas our scheme achieves a slightly smaller rate $2/5$ using $u=1$. 
\end{remark*}

\begin{example}
\label{example:stargraphPIRk=9}
\begin{figure}[t]
\centering
\resizebox{0.25\columnwidth}{!}{%
\begin{tikzpicture}[scale=0.8,
    every node/.style={font=\tiny},
    every edge/.style={font=\tiny}]

  \node[circle, draw, fill=blue!20, minimum size=3mm] (c) at (0,0) {$10$};

  % leaves
  \foreach \i/\ang in {1/90,2/50,3/10,4/-30,5/-70,6/-110,7/-150,8/-190,9/-230} {
    \node[circle, draw, fill=red!20, minimum size=3mm] (n\i)
      at ({1.35*cos(\ang)}, {1.35*sin(\ang)}) {$\i$};
  }

  % edges (above)
  \foreach \i in {9,8,7,6} {
    \draw (c) -- (n\i) node[midway, sloped, above, scale=0.7] {$W_{\i}$};
  }

  % edges (below)
  \foreach \i in {1,2,3,4,5} {
    \draw (c) -- (n\i) node[midway, sloped, below, scale=0.7] {$W_{\i}$};
  }

\end{tikzpicture}}
\caption{\small Star graph with $K=9$ spoke vertices and the central hub vertex, sharing $9$ files.}
\label{figure:stargraph}
\end{figure}
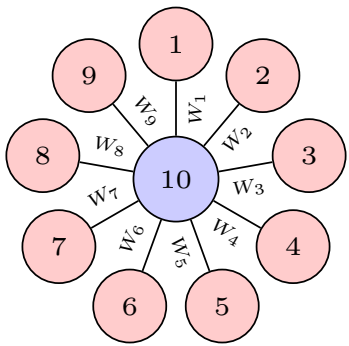

Consider a star network with a central hub and $K=9$ spoke servers, $S_1, \dots, S_9$, each storing a file, as shown in Fig. \ref{figure:stargraph}. Suppose the desired file is $W_1$. We choose $u=2$, as $(u+1)|K$. We assume $L=1$ bit. The protocol is executed as follows.
%%%
\begin{enumerate}[leftmargin=*]
    \item The client randomly selects a set $U$ of $u=2$ server indices, say $U=\{5, 6\}$, and requests files $W_5$ and $W_6$ from the corresponding spoke servers.
    \item Since the desired file index $1 \notin U$, the client must query the hub. It constructs a query matrix $M$ of size $(u+1) \times \frac{K}{u+1} = 3 \times 3$. The matrix is a random permutation of $\{1, \dots, 9\}$, with the crucial constraint that the desired index $\theta=1$ is placed in the same column as the elements of $U=\{5,6\}$. A possible query matrix is:
    \[
    M = \begin{bmatrix}
        3 & \mathbf{5} & 7 \\
        2 & \mathbf{6} & 4 \\
        9 & \mathbf{1} & 8
    \end{bmatrix}
    \]
    \item The hub responds with the XOR sum of files for each column.
    \item  Using the second column-sum, $A_N^2 = W_5 \oplus W_6 \oplus W_1$ and the already retrieved files $W_5$ and $W_6$, the client decodes its desired file: $W_1 = A_N^2 \oplus W_5 \oplus W_6$.
\end{enumerate}
Note that, if the initial random set had contained the desired index (e.g., $U=\{1,5\}$), the hub would not have been queried at all as $W_1$ would be retrieved already. 
\end{example}
%%%%
We now provide an informal argument for privacy (the formal protocol and proof follow later in the section). Note that each spoke server $S_i$ is queried with probability $u/K = 2/9$, regardless of the desired file $\theta$. On the other hand, the hub either receives no query (with probability $2/9$) or a matrix where $\theta$ is hidden in a random position, making it statistically indistinguishable from any other file index.

 We now calculate the rate. With probability $u/K=2/9$, the total download is $u=2$ bits (when the file is retrieved directly from the spoke servers). Otherwise, With probability $(K-u)/K=7/9$, the total download is $u + \frac{K}{u+1} = 2+3=5$ bits. Thus, expected download is $D = (2/9)\cdot 2 + (7/9)\cdot 5 = 13/3$ bits. The rate is thus $R = 1/D = 3/13$.

Now, we begin to formally describe the protocol. Throughout, we assume the subpacketization is $L=1$ bit. 
%%%
\subsection{Query generation}
Suppose the desired file index is $\theta \in [K]$. 
The client chooses %\pk{You wrote `Choose'. I changed it a bit to reflect what wil happen. You can write in similar language}
a non-negative integer $u$ such that $(u+1)$ divides $K$. 

Later, we will optimize the value of $u$ such that it minimizes the  expected download, given by $\frac{u^2 + K}{u+1}$ (this is shown in Subsection \ref{subsec:rate}). 

The client then chooses a $u$-sized subset $U$ uniformly at random from the set of all $u$-sized subsets of file-indices $[K]$. For each $i\in U$, i.e., the client then queries the server $S_i$ for the file $W_i$, i.e., the following queries are sent to the spoke servers: 
\begin{align*}
    Q_i = \begin{cases}
        i \text{ if } i \in U \\
        \phi \text{ if } i \notin U
    \end{cases} \forall i \in \{1,2,\dots, N-1\}. 
\end{align*}
Here $\phi$ denotes the event that no query is sent.  

Now, if $\theta \in U$, then the hub server is not queried, i.e., $Q_N=\phi$, and the protocol completes. Otherwise, i.e. if $\theta \notin U$, the query to the central server is generated as follows.

\subsubsection*{Query to hub server if $\theta\notin U$} Let $a = \frac{K}{u+1}$. Let $\mathcal{M}$ denote the set of all $(u+1) \times a$ matrices that contain each element of $[K]$ appearing exactly once. The client chooses a random \textit{query matrix} $M$ from ${\cal M}$, as per following steps. 

\begin{enumerate}[leftmargin=*]
\item Uniformly at random choose a row index $R \in [u+1]$. Uniformly at random choose a column index $C \in [a]$. The entry in the row-$R$ and column-$C$ is fixed as $\theta$, i.e., $M(R,C)=\theta$ 
% \pk{I changed the random $i,j$ to $R,C$ because I feel $i,j$ can (and ARE) overused. Suitably editing other places. But do check} 
\item A permutation $\sigma_U$ of $U=\{k_{1},\hdots,k_{u}\}$ is chosen uniformly at random from the set of $u!$ such permutations. This permutation determines the ordering of the elements from $U$ to be placed in column $C$. These elements populate the available rows $\{1, \dots, u+1\} \setminus \{R\}$ sequentially. Specifically, the $i^{\text{th}}$ element of the permuted sequence, $\sigma_U(k_i)$, is placed in the $i^{\text{th}}$ available row index, for $i=1, \dots, u$.
% \pk{Vayur, Check this and the next point}
\item Next, the set of remaining values, $V = [K]\setminus (U\cup\{\theta\})$, is used to fill the rest of the matrix. A permutation $\sigma_V$ of the elements in $V$ is selected uniformly at random from the collection of $(K-u-1)!$ such permutations. The elements of $V$ are then permuted as per $\sigma_V$ and then placed into the $(a-1)(u+1)$ unfilled positions in $M$, following a canonical sequence (say, filling all empty cells in column 1 from top to bottom, then all empty cells in column 2, and so on, except for the $C^{\text{th}}$ column, as per the permutation $\sigma_V$).

\end{enumerate}
%%%%
With this construction of $M$, we send $Q_N = M$ to the central server (if $\theta \notin U$).

\subsection{Responses}
At any server, if the query $Q_i=\phi$, that means no query is essentially seen, which implies no response is sent. At any spoke servers $S_i:i\in[N-1]$, if $Q_i = i$, the server returns the file $W_i$ stored at server $S_i$. 

As for the hub server, if $Q_N = M$, the server is expected to return $a$ bits, corresponding to the XOR of $(u+1)$ files for each column of $M$. Specifically, the hub server responds with the vector $A_N = (A_N^1,A_N^2,\dots,A_N^a)$, where $A_N^l:=\bigoplus\limits_{m = 1}^{u+1} W_{M(m,l)}$. 

\subsection{Verifying Correctness}
For the case where $\theta\in U$, the client receives $W_\theta$ from the $\theta$-th spoke server, and retrieval is trivial.
For the case where $\theta\notin U$, suppose the client had chosen $C=c$, to construct the matrix $M$. Consider the $c^{\text{th}}$ coordinate of the answer from the central server, given by
$A_N^c= \bigoplus_{m = 1}^{u+1} W_{M(m,c)}$.
From the spoke servers, we have files 
$W_{k}, \forall k \in U$
Note that 
$\bigcup_{m=1}^{u+1} M(m,c) = U \cup \theta$
Therefore, we can XOR the files $W_k:k\in U$ with $A_N^c$ to obtain $W_\theta$, i.e., the client obtains.
$$
W_\theta = (\bigoplus_{k \in U} W_k) \oplus A_{N}^c.
$$
This completes the proof of correctness of the protocol. 
\subsection{Verifying Privacy}
To show that the protocol is private, we must show that the distribution of queries $P_{Q_i}$ is independent of $\theta$ for all the servers. 

For each spoke server $i \in [N-1]$, we have ${\cal Q}_i=\{\phi,i\}$. Thus, by our random choice of $U$, we have 
\begin{align}
\label{eqn:distriQueries_spoke}
    P_{Q_i|\theta}(i|k) &= P(i \in U) = \frac{\binom{K-1}{u-1}}{\binom{K}{u}}= \frac{u}{K}= 1- P_{Q_i|\theta}(\phi|k).
\end{align}

For the central server $S_N$, we have ${\cal Q}_N=\{\phi\}\cup{\cal M}$. We first observe that 
\begin{align}
\label{eqn:distriQueries_hub1}
    P_{Q_N|\theta}(\phi|k)
    = P(\theta \in U)
    = \frac{\binom{K-1}{u-1}}{\binom{K}{u}}
    = \frac{u}{K}. 
\end{align}
% Hence probability of seeing the null query at the hub server is independent of $\theta$.
For some specific $M_p\in{\cal M}$, we now calculate $P_{Q_N|\theta}(M_p|k)$. Suppose that $M_p(s,t)=k$. Further, let $\sigma_1$ be the ordered set of the $u$ elements (denoted by $U_1$) apart from $k$ in column-$t$, and $\sigma_2$ be the ordered set of the $(a-1)(u+1)$ elements in the other columns apart from column-$t$ (according to the canonical order). Thus, for the event $M=M_p$ to occur given $\theta=k$, it must be that $(R,C)=(s,t)$, $U=U_1$, $\sigma_U=\sigma_1$ and $\sigma_V=\sigma_2$. Thus, 
\begin{align}
\nonumber
P_{Q_N|\theta}(M_p|k)&=P((R,C)=(s,t))P_U(U_1)P_{\sigma_U}(\sigma_1)P_{\sigma_V}(\sigma_2)\\
\nonumber
    &= \frac{1}{(u+1)a}\cdot  \frac{1}{\binom{K}{u}} \cdot \frac{1}{u!} \cdot \frac{1}{(K-u-1)!} \\
\label{eqn:distriQueries_hub2}
    &=  \left(\frac{K-u}{K}\right)\frac{1}{K!} 
\end{align}

From \eqref{eqn:distriQueries_spoke},\eqref{eqn:distriQueries_hub1}, and \eqref{eqn:distriQueries_hub2}, we see that the distribution $P_{Q_i|\theta}$ is independent of $\theta$. We thus conclude that the protocol is private. 
% probability of receiving $M_p$ does not depend on $\theta$.combining (a), (b), (c), (d), (e), We see that $P(Q_i = q_i) \perp\!\!\! \text{ }\text{ } \theta \forall q_i \in \mathcal{Q}_i \forall i \in [N]$. Following the logic presented in~\eqref{eq:independence} we conclude that the protocol is private.
\subsection{Rate} \label{subsec:rate}
The rate of a protocol can be calculated as $R = \frac{L}{D}$, where $L$ is the subpacketization (measured in bits) and $D$ is the expected download.  
% We take  $L = 1$.
To calculate expected download, we note that we download one bit from $u$ spoke servers, and download $a$ bits from the hub server if $(\theta \notin U)$.
Therefore, denoting our download as $D(u)$, we get 
\begin{align*}
    D(u) &= u + P(\theta \notin U)a 
    = u + \frac{K-u}{K} \cdot \frac{K}{u+1}= \frac{u^2 + K}{u+1}. 
\end{align*}
Thus, the rate of the protocol is 
\begin{align}
   R(u) = \frac{u+1}{u^2 + K}. \label{eq:rateofstarPIR_probabilistic}
\end{align}
%%%%%
\subsection{Optimizing $u$ and the role of dummy files}
Note that, to execute our scheme, we may use any value of $u$ satisfying the constraint $(u+1)|K$. For a given $K$, it remains to find the optimal integer $u^*\in[1:K]$ such that $D(u^*)\leq D(u)$, under the constraints ${K}/{(u+1)}\in \mathbb{Z}$. Without the integral constraint, we can obtain the optimal solution  $u^*=\sqrt{K+1}-1$, as this is the minima of $D(u)$ under the constraints. 

This satisfies our scenario if $N=K+1$ is a perfect square. Otherwise, we add $K'-K$ dummy files (along with dummy spoke servers), where $K'$ is the smallest integer such that $K'\geq K$ and $K' + 1$ is a perfect square. We can then obtain $u^*=\sqrt{K'+ 1}$ as our optimal choice. The corresponding scheme on $K'$ files (along with the $K'-K$ dummy files) is clearly still private and correct. Observe that  $\lceil\sqrt{K+1}\rceil = \sqrt{K'+1} $ and $K' + 1 < \left(\sqrt{K+1} + 1 \right)^2= K + 2 + 2\sqrt{K+1}$. Thus, we need to add at most $2\sqrt{K+1}+1$ dummy files\footnote{Upto $K = 810000$, we have run a simulation to verify that at most $58$ dummy files are required to keep the download optimal}. 

Thus, we see that our protocol achieves the download $D(u^*)=2\sqrt{K'+ 1}-2+\frac{1}{\sqrt{K'+ 1}+ 1}$, the rate achieved being its reciprocal. Along with the fact that $K=N-1$, this proves Theorem \ref{thm:starvar}.

%%%%%%%%%%%%%%%%%%%%%%%%%%%%%%%%%%%%%%%%%%%%%%%%%%%%%%%%%%%%
\section{A New PIR scheme for general graphs with $L=1$}
\label{sec:generalgraph1}

In this section, we show a general variable-download PIR scheme for any graph. Our scheme uses a clever recursive query-construction strategy, based on independent sets of the given graph. Our scheme uses subpacketization $L=1$. The rate achieved by this scheme is characterized by the following theorem and corollary. 
%%%%
\begin{theorem} \label{thm:gengraphs}
    There exists a variable-download PIR scheme for general graphs over $N$ vertices that employs subpacketization $L = 1$ and achieves a rate $R \geq \max\left(\frac{2}{2N - \alpha(G)},\frac{1}{N-1}\right)$, where $\alpha(G)$ is the independence number of the graph.
\end{theorem}
Theorem \ref{thm:gengraphs} leads to the following corollary immediately, by observing that the set of vertices on any side of the bipartite is an independent set. 
%%%%
\begin{corollary}
    Our scheme with subpacketization $L=1$ yields rate $\frac{1}{N-1}$  for the complete graph $K_N$, and thus for any graph. Further, for any bipartite graph with $N_1\geq N/2$ vertices on one-side, we obtain $R=\frac{2}{2N-N_1} \geq \frac{4}{3N}$. 
\end{corollary}
%%%%
\begin{remark*}[Comparison with prior work]
For complete graphs, the rates achieved in prior work is summarized are Table~\ref{table:pir_graphs_previousWork}. The rates of prior schemes are $\tfrac{2^{N-1}}{2^{N-1}-1}\tfrac{1}{N}$, $\tfrac{6}{5-2^{3-N}}\tfrac{1}{N}$ and $\approx (1-o(1))\tfrac{4}{3N}$
in \cite{Sadehetal_TIT_2023_BoundsonPIRgraphs, Kong_TIT_2025newcapacityboundspir, ge2025privateinformationretrievalgraphs} respectively. Although these schemes have an inherently high subpacketization, they can be converted into variable-download, unit-subpacketization schemes employing \cite[Theorem 1.14]{ge2025privateinformationretrievalgraphs}. All of these schemes, including ours are $\Theta(\frac{1}{N})$. However, the rate achieved by Theorem \ref{thm:gengraphs} is smaller than these prior ones. The tightest capacity upper bound known \cite{ge2025privateinformationretrievalgraphs} is $\approx \frac{1.3922}{N}$. The benefit of our scheme is that size of the query set is small, thereby implying a lower upload cost.

Another scheme specialized for sparse graphs is mentioned in \cite[Theorem 1.15]{ge2025privateinformationretrievalgraphs}, that achieves $R = \frac{1}{N - \sum_{n = 1}^N2^{-d_n}}$. Our scheme also does increasingly well as the sparsity increases, obtaining  a better rate than \cite[Theorem 1.15]{ge2025privateinformationretrievalgraphs} in all cases. 

For complete bipartite graphs, our scheme achieves the best known rate in literature. Note that the upper bound for bipartite graphs is $\approx \frac{1.5415}{N}$~\cite{ge2025privateinformationretrievalgraphs}, whereas our scheme achieves $R \approx \frac{1.333}{N}$.
\end{remark*}

The rest of this Section focuses on proving Theorem \ref{thm:gengraphs}. We present the description of the scheme, prove its privacy and correctness, and finally obtain the rate. 
%%%
\subsection{Preliminaries} 
% Consider the PIR problem for $N$ servers and $K$ files, where the storage configuration is described by a graph.
Consider the PIR graph $G(V, E)$ with vertices set $V=[N]$, representing the servers. Let the degree of the $n$-th vertex be denoted by $d_n$, i.e., $d_n$ files are present are server $n$. Apart from the standard notation for the files $\{W_{i,j}:\{i,j\}\in E\}$, we also represent equivalently the files at server $n$ as $W_1^n, W_2^n, \dots W_{d_n}^n$. 

We assume that the vertex set $V$ is partitioned into $\kappa$ disjoint independent sets $I_1,\hdots,I_{\kappa}$, satisfying the below property.
\begin{itemize}[leftmargin=*]
    \item The subset $I_i$ is a largest independent set of the graph $G[I_i \cup \dots \cup I_{\kappa}]$. 
\end{itemize}
\begin{remark*}
For the PIR protocol presented in this section, it is sufficient for $I_i:i\in[\kappa]$ to be maximal, not necessarily the largest. Note that, while finding the largest independent set is computationally hard, we can construct a maximal independent set of a graph  in polynomial-time, using a simple greedy algorithm.
\end{remark*}

We now begin with the description of the protocol. An example for the same is provided in Example \ref{subsec:example-gengraph} that appears subsequently. 
%%%%
\subsection{General structure of queries received, and the responses}
For the $n$-th server with degree $d_n$, the query vector constructed by the client is a $d_n$-length vector $Q_n= (f_1^n, f_2^n, \dots, f_{d_n}^n)$, 
% \begin{align}
%     q_n = (f_1^n, f_2^n, \dots, f_{d_n}^n) \label{eq:query}
% \end{align}
where $f_j^n \in \Fb_2, \forall j$. We assume that the event $Q_n = \overline{0}$ corresponds to the empty-query, for which no response is sent by server $n$. 

Upon receiving a query $Q_n \neq \overline{0}\in \Fb_2^{d_n}$, the server responds with the linear combination of the files it has, weighted by the query it received. Hence, the response is 
    $A_n = \oplus_{j=1}^{d_n}f_j^nW_j^n$.
\subsection{How queries are generated}
We now describe the process of sampling the random query vectors $Q_n:n\in[N]$. Note that, we have 
$V = I_1 \sqcup I_2 \sqcup \dots \sqcup I_{\kappa}$. Further, no two vertices in any specific independent set $I_i:i\in[\kappa]$ share a file. Suppose the desired file is $\theta = W_{n,m}$ with $n \in I_x, m\in I_y \text{  , } x \neq y$. Hence $\theta = W_{n,m} = W_a^n = W_b^m $ for arbitrary but fixed $a,b$.

The client generates queries sequentially, first for $I_1$, then $I_2$ and so on. We describe this process in a sequential manner, similarly. 
\subsubsection{Step 1: Queries for servers in $I_1$}
For each $n \in I_1$, the client flips a coin and with equal probability assigns 
\[
Q_n = \begin{cases}
    \overline{1}\in \Fb_2^{d_n},& \text{ with probability } \frac{1}{2} \\
    \overline{0}\in \Fb_2^{d_n},& \text{ with probability } \frac{1}{2}
\end{cases}
\]
where $\overline{1}$ denotes the all ones vector and $\overline{0}$ denotes the all zeros vector.

We now show the $s^{\text{th}}$ step in the process for $I_s$, for each $2\leq s\leq \kappa-1$. 

\subsubsection{Step s: Queries for servers in $I_s$, for $s\in[2:\kappa]$} We describe the query generation for server $n$, for each $n\in I_s$. We can partition the edges incident at $n$ into two categories, namely 
\begin{itemize}[leftmargin=*]
    \item The \textit{downstream} edges with the other end point at $m \in (I_{s+1} \cup \dots \cup I_{\kappa})$. Observe that when $s=\kappa$, there are no downstream edges surely.
    \item The \textit{upstream} edges with the other end point at $m \in I_1 \cup \dots I_{s-1}$. Note that at least one such edge exists, as otherwise $n$ could have been added to any of $I_1,\dots,I_{s-1}$, thereby contradicting their maximality.
\end{itemize}
We treat the two cases differently. Let $d^{\mathsf{\downarrow}}_n$ be the degree of $n$ in the induced subgraph  $G[I_s \cup \dots I_{\kappa}]$, i.e. the number of downstream edges. We denote the files corresponding to the downstream edges as $(W_1^n, W_2^n \dots W_{d_n^\downarrow}^n)$. Let the files corresponding to the upstream edges at vertex $n$ be $(W_{d_n^\downarrow + 1}^n, \dots W_{d_n}^n)$. Corresponding to the downstream edges (if $d_n^\downarrow\neq 0$), we assign the \textit{downstream query subvector} of $Q_n$ as
\[
(f_1^n, f_2^n, \dots f_{d_n^{\downarrow}}^n) = \begin{cases}
    \overline{1}\in\Fb_2^{d_n^\downarrow}, \text{ wp } \frac{1}{2} \\
    \overline{0}\in \Fb_2^{d_n^\downarrow}, \text{ wp } \frac{1}{2},
\end{cases}
\]
where $\text{wp}$ denotes `with probability'. 

The \textit{upstream query subvector}, corresponding to the upstream edges at $n$, is determined by the weights assigned to the upstream files when constructing the query for the corresponding upstream servers. To be precise, for each $j \in [d_n^\downarrow+1 : d_n]$, we set  $f_j^n$ as follows: 
\begin{align*}
f_j^n &= 
\begin{cases}
    f_a^m & \text{if } W_\theta \neq W_{n,m} \\
    f_a^m \oplus 1 & \text{if } W_\theta = W_{n,m},
\end{cases}
\end{align*}
where $f_a^m$ is the weight assigned to file $W_{n,m}$ while querying $m$.
The client finally concatenates the upstream and the downstream query subvectors to construct the final query vector $Q_n=(f_1^n, f_2^n, \dots, f_{d_n}^n)\in \Fb_2^{d_n}$.

\subsection{Correctness}\label{subsec:gengraph_correctness}
The query design ensures that when all server responses are XORed together, every undesired file cancels out, leaving only the desired file $W_\theta$. Consider an arbitrary file $W_{n,m}$ stored on servers $n$ and $m$, with respective query bits $f^n$ and $f^m$. The total contribution of this file to the final sum (of all responses) is $(f^n \oplus f^m)W_{n,m}$. If $W_{n,m}$ is not the desired file ($W_{n,m} \neq W_\theta$), the query construction ensures the weights are identical ($f^n = f^m$), whereas if it is the desired file ($W_{n,m} = W_\theta$), the weights are different ($f^n \neq f^m$). Therefore, the coefficient $(f^n \oplus f^m)$ is 0 for all undesired files and 1 for the desired file. Consequently, the sum over all server responses results in $W_\theta$, i.e., we obtain
\begin{align*}
    \bigoplus_{i=1}^{N} A_i = W_\theta
\end{align*}

%%%%%
\subsection{Proof of Privacy and the Distribution of the Queries}
The formal proof of privacy, detailed in the Appendix \ref{app:ProofofPrivacy_PIRgeneralgraph}, proceeds by induction on the sequence of independent sets $I_1, \dots, I_{\kappa}$. The summary of the proof is provided here. 
\begin{enumerate}[leftmargin=*]
    \item \textit{Base Case:} For servers in the first set, $I_1$, privacy holds by construction. 
    % Their queries are chosen to be the all-zeros or all-ones vector with equal probability, a process that does not depend on $W_\theta$.

    \item \textit{Inductive Step:} We assume that queries to all servers in the preceding sets ($I_1, \dots, I_{s-1}$) are private. For a server $n \in I_s$, its query vector $Q_n$ is constructed from two subvectors, the downstream subvector $Q_n^{\downarrow}$ and the upstream subvector $Q_n^{\uparrow}$. We note the following. 
    \begin{itemize}[leftmargin=*]
        \item By construction, $Q_n^{\downarrow}$ is independent of $\theta$.
        \item The upstream subvector $Q_n^{\uparrow}$ is deterministically calculated using query bits of vertices in prior independent sets. Thus, by the inductive hypothesis, the query bits of these upstream subvector are already independent of $\theta$. Further, by construction these are also independent of $Q_n^{\downarrow}$, given $\theta$. Also, the protocol is designed such that the individual upstream query bits of $Q_n^{\uparrow}$ are themselves pairwise independent and uniformly random. 
    \end{itemize}
\end{enumerate}
Using these arguments that are elaborated in Appendix \ref{app:ProofofPrivacy_PIRgeneralgraph}, we can obtain the distribution of $Q_n$ given $\theta$ explicitly as follows. 

Let $q = (q^{\downarrow}, q^{\uparrow})\in \Fb_2^{d_n}$ be a potential query vector for server $n$, where $q^{\downarrow}$ and $q^{\uparrow}$ correspond to the downstream and upstream subvectors respectively. Note that we have the following to be true by construction. 
$$
\text{If}~d_n^\downarrow\neq 0,P(Q_n^{\downarrow} = q^{\downarrow} \mid \theta) = \begin{cases} 1/2 & \text{if } q^{\downarrow} \in \{\overline{0}, \overline{1}\} \\ 0 & \text{otherwise} \end{cases}. 
$$
If $d_n^\downarrow= 0$, $Q^\downarrow$ is undefined. In this case, we assign $P(Q_n^{\downarrow} = q^{\downarrow} \mid \theta) = 1$ for convenience. As for the upstream subvector, by prior arguments, we have 
$ P(Q_n^{\uparrow} = q^{\uparrow} \mid \theta) = \left(\frac{1}{2}\right)^{d_n - d_n^{\downarrow}}. $
By the induction arguments prior, we can thus compute the distribution of $Q_n$ given $\theta$, as follows. 
\begin{align*}
P&(Q_n = q \mid \theta) = P(Q_n^{\downarrow} = q^{\downarrow}, Q_n^{\uparrow} = q^{\uparrow} \mid \theta) \\
&= P(Q_n^{\downarrow} = q^{\downarrow} \mid \theta) P(Q_n^{\uparrow} = q^{\uparrow} \mid \theta)\\
&= \begin{cases} \left(\frac{1}{2}\right)^{d_n - d_n^{\downarrow}+1}, & \text{if } d_n^\downarrow\neq 0 \text{~and~} q^{\downarrow} \in \{\overline{0}, \overline{1}\}\\ \left(\frac{1}{2}\right)^{d_n - d_n^{\downarrow}},& \text{if } d_n^\downarrow= 0\\ 0, & \text{otherwise}. \end{cases}
\end{align*}
We thus confirm that the query $Q_n$ is independent of $\theta$. This completes the induction, proving that the protocol is private at all servers.

\subsection{Rate}
We have subpacketization $L = 1$.
Note that for almost all queries, we receive one bit from each server, except for the times when 
$Q_n = \overline{0}\in\Fb_2^{d_n}$ in which case we download zero bits. Thus, the total expected download is 
    % D = \sum_{n \in V} \left(1 - P(q_n = \bar{0})\right)
\begin{align}
\nonumber
D & = \sum_{n \in V} \left(1 - P(Q_n = \bar{0})\right)\\
\label{expr:download}
&\stackrel{(a)}{\leq} N -\sum_{n \in I_1} \tfrac{1}{2}-\sum_{s=2}^{\kappa-1}\sum_{n\in I_s}\left(\tfrac{1}{2}\right)^{d_n - d_n^{\downarrow}+1}-\sum_{n\in I_{\kappa}}\left(\tfrac{1}{2}\right)^{d_n - d_n^{\downarrow}}\\
\label{eqn:upperboundDownload_gengraph1}
&\stackrel{(b)}{\leq}N -\frac{\alpha(G)}{2}, 
\end{align}
where the inequality $(a)$ is because $d_n^\downarrow$ could be $0$ for some $n\in I_s, s\in[2:\kappa-1]$, and $(b)$ follows since $\alpha(G)=|I_1|$ as $I_1$ is defined as a largest independent set of $G$. 

Now, the term $N-\alpha(G)/2\leq N-1$ provided $\alpha(G)\geq 2$. Otherwise, $\alpha(G)=1$, which implies that $G$ must be complete. For the complete graph $K_N$, our partition of the $N$ vertices must precisely be $I_1\cup I_2\cup\dots\cup I_N$, where $|I_i|=1$ (each independent set must be a single vertex). Furthermore, by observation of the graph structure, we can say that $d_n-d_n^\downarrow=s-1,$ for $n\in I_s$ (there are exactly $s-1$ upstream edges for the vertex in $I_s$). Thus, the download in this case is given as 
\begin{align}
    D=N-\sum_{s=1}^{N-1}\frac{1}{2^s}-\frac{1}{2^{N-1}}=N-1. 
\end{align}

As the rate $R = \frac{1}{D}$, we see that 
this completes the proof of Theorem \ref{thm:gengraphs}.
%%%

\subsection{Extension to $r$-multigraphs}
\label{subsec:multigraph}
After completing the proof of Theorem~\ref{thm:gengraphs}, we are in a position to discuss the extension of the scheme to $r$-multigraphs $G^{(r)}$, where each edge in $G$ is replaced by $r$ parallel edges.

\begin{corollary} \label{cor:multigraph}
    The scheme in Theorem~\ref{thm:gengraphs} over $G$ can be extended to $G^{(r)}$ with subpacketization $L = 1$, achieving an expected rate:
    \begin{equation}
        R \geq \max\left(\frac{1}{N - \alpha(G)2^{-r}}, \frac{1}{N-2^{1-r}}\right)
    \end{equation}
    %where for the complete multigraph $K_N^{(r)}$, the rate is $R_{\text{comp}}^{(r)} = \left(N - \sum_{s=1}^N 2^{-sr}\right)^{-1} \approx \frac{1}{N - \frac{1}{2^r - 1}}$.
\end{corollary}

\begin{IEEEproof}
    We modify the construction of the downstream query subvector $Q_n^\downarrow$. In $G^{(r)}$, each vertex $n$ has $r$ parallel edges for every adjacent downstream server. We partition these edges into $r$ groups, where each group contains one edge per parallel set. The client samples $r$ independent Bernoulli variables $(X_{n,1}, \dots, X_{n,r})$ and assigns $X_{n,k}$ as the weight for all downstream edges in group $k$. 
    
    The upstream subvector is determined as in the simple-graph case: for the $k$-th parallel edge between $n$ and an upstream server $m$, the weight $f_{n,m,k}$ is set to $f_{m,n,k} \oplus \mathbb{I}(W_{n,m,k} = W_\theta)$. Privacy is maintained because the $r$ Bernoulli variables are mutually independent and independent of $\theta$. For the complete multigraph, (assuming that the independent sets are $I_j=\{j\}:j\in[N]$) the probability that server $s \in [N]$ receives a null query is $P(Q_s = \overline{0}) =  \begin{cases} 2^{-sr} \text{,} \quad s \in V\setminus I_N \\2^{-(N-1)r} \text{,} \quad s \in I_N\end{cases}$,
    The expected download is thus $D = N - 2^{1-r}$. If the graph is not complete, then $\alpha({G}) \geq 2 $ and $D \geq N - \alpha(G)2^{-r}$, proving the corollary. 
\end{IEEEproof}
The rate achieved by Corollary~\ref{cor:multigraph} for the multi-graph extension $K_N^{(r)}$ of the complete graph is $R = \frac{1}{N-2^{1-r}}$, the best known in the literature (see Table \ref{table:pir_graphs_previousWork}, for the prior bounds from \cite{Kong_TIT_2025newcapacityboundspir, ge2025privateinformationretrievalgraphs}). Note that the known upper bound \cite{Kong_TIT_2025newcapacityboundspir} on the capacity is $C(K_N^{(r)}) \leq \frac{1}{N - (N-1)2^{-r}}$, and our rate expression has a similar structure.

\begin{example}
\label{subsec:example-gengraph}
\begin{figure}[t]
\centering
\begin{tikzpicture}[scale=0.85,
    every edge/.style={draw},
    vertex/.style={circle, draw, minimum size=8mm, font=\normalsize},
    edgelabel/.style={font=\scriptsize, inner sep=0pt}]

  % existing vertices
  \node[vertex, fill=orange!65!white] (1) at (0,0)   {$1$};
  \node[vertex, fill=green!45!white]  (2) at (2,0)   {$2$};
  \node[vertex, fill=blue!55!white]   (3) at (1,1.6) {$3$};
  \node[vertex, fill=orange!65!white] (4) at (3,1.6) {$4$};
  \node[vertex, fill=blue!55!white]   (5) at (4,0)   {$5$};

  % new vertices
  \node[vertex, fill=green!45!white]  (6) at (6,0)   {$6$};
  \node[vertex, fill=green!45!white]   (7) at (5,1.6) {$7$};

  % existing edges
  \draw (1)-- node[midway, sloped, below, edgelabel] {$W_{1,2}$} (2);
  \draw (1)-- node[midway, sloped, above, edgelabel] {$W_{1,3}$} (3);
  \draw (2)-- node[midway, sloped, above, edgelabel] {$W_{2,3}$} (3);
  \draw (2)-- node[midway, sloped, above, edgelabel] {$W_{2,4}$} (4);
  \draw (3)-- node[midway, sloped, above, edgelabel] {$W_{3,4}$} (4);
  \draw (4)-- node[midway, sloped, above, edgelabel] {$W_{4,5}$} (5);

  % new edges
  \draw (5)-- node[midway, sloped, below, edgelabel] {$W_{5,6}$} (6);
  \draw (4)-- node[midway, sloped, above, edgelabel] {$W_{4,7}$} (7);
  \draw (5)-- node[midway, sloped, above, edgelabel] {$W_{5,7}$} (7);

\end{tikzpicture}
\caption{PIR Graph for problem on $7$ servers and $9$ files. A scheme for this graph is presented in Example \ref{subsec:example-gengraph}.}
\label{fig:general_graph_7}
\end{figure}
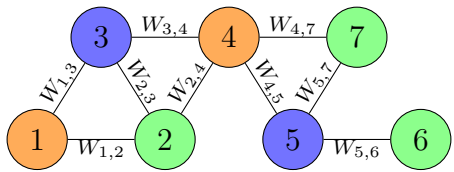

We now give an example to help illustrate our scheme for the simple graph. Consider the PIR graph described by Figure~\ref{fig:general_graph_7} representing a PIR problem on $N=7$ servers. Suppose we want to retrieve $W_\theta = W_{2,3}$

We partition the vertices of the graph into three independent sets based on the coloring shown in Fig. \ref{fig:general_graph_7}: $I_1 = \{2, 6, 7\}$ (green), $I_2 = \{1, 4\}$ (orange), and $I_3 = \{3, 5\}$ (blue). The client uses independent Bernoulli random variables $X_i, Y_i \sim \text{Bern}(1/2)$ to construct the queries $Q_n:n\in[N]$ to the $N$ servers. The specific query sent during an instantiation of the protocol is obtained by sampling the variables $X_i,Y_i$. 

In Step 1, for each server $n\in I_1$, the client assigns identical weights to all incident \textit{downstream} edges (incident on $n$ and on some vertex in $I_2\cup I_3$) using a single random variable $X_i$. Specifically, for server-$2$, the client fixes the query $Q_2 = (f_{2,1}^2, f_{2,3}^2, f_{2,4}^2) = (X_2, X_2, X_2)$. This query vector specifies the linear combination to downloaded from the server as a response. That is, the server-$2$ sends $A_2=f_{2,1}^2W_{2,1}\oplus f_{2,3}^2W_{2,3}\oplus f_{2,4}^2W_{2,4}$ as a response. Similarly, the client sets $Q_6 = (f_{6,5}^6) = (X_6)$, and $Q_7 = (f_{7,4}^7, f_{7,5}^7) = (X_7, X_7)$. 

In Step 2, the client constructs queries for servers in $I_2$ by matching the weights of upstream edges to those assigned in Step 1 and sampling new variables for downstream edges. For server-$1$, the weight of the upstream edge $(1,2)$ is $f_{1,2}^1 = f_{2,1}^2 = X_2$, and the downstream edge $(1,3)$ is assigned $Y_1$, yielding $Q_1 = (f_{1,3}^1, f_{1,2}^1) = (Y_1, X_2)$. Similarly, for Server 4, the upstream weights $f_{4,2}^4$ and $f_{4,7}^4$ match $X_2$ and $X_7$ respectively, while downstream edges $(4,3)$ and $(4,5)$ are assigned $Y_4$, resulting in $Q_4 = (f_{4,3}^4, f_{4,5}^4, f_{4,2}^4, f_{4,7}^4) = (Y_4, Y_4, X_2, X_7)$.

In Step 3, queries for $I_3$ are formed by matching all incident upstream edges to weights from $I_1$ and $I_2$. To retrieve the desired file $\theta = W_{2,3}$, we apply the shift $f_{3,2}^3 = f_{2,3}^2 \oplus 1$. Thus, Server 3 receives $Q_3 = (f_{3,1}^3, f_{3,2}^3, f_{3,4}^3) = (Y_1, X_2 \oplus 1, Y_4)$, and Server 5 receives $Q_5 = (f_{5,4}^5, f_{5,6}^5, f_{5,7}^5) = (Y_4, X_6, X_7)$. The responses $A_j:j\in[N]$

\textit{Correctness:} The client computes the XOR sum of all responses $\bigoplus_{n=1}^7 A_n$. For any edge $(u,v) \neq (2,3)$, the weights $f_{u,v}^u$ and $f_{v,u}^v$ are identical by construction (e.g., $f_{1,3}^1 = f_{3,1}^3 = Y_1$), causing these files to cancel out. For the desired file, the weights are $f_{2,3}^2 = X_2$ and $f_{3,2}^3 = X_2 \oplus 1$. Their XOR sum is $X_2 \oplus X_2 \oplus 1 = 1$, leaving exactly $W_{2,3}$.

\textit{Expected Rate and Privacy:} The download cost is determined by the probability that each query vector $Q_n \neq \overline{0}$. For $I_1$, $P(Q_2 \neq \overline{0}) = P(Q_6 \neq \overline{0}) = P(Q_7 \neq \overline{0}) = 1/2$. For $I_2$, $P(Q_1 \neq \overline{0}) = 1 - (1/2)^2 = 3/4$ and $P(Q_4 \neq \overline{0}) = 1 - (1/2)^3 = 7/8$. For $I_3$, $P(Q_3 \neq \overline{0}) = 1 - P(Y_1=0, X_2=1, Y_4=0) = 7/8$ and $P(Q_5 \neq \overline{0}) = 1 - (1/2)^3 = 7/8$. The total expected download is $D = \sum_{n\in [N]}P(Q_n\neq \overline{0})=1.5 + 0.75 + 2.625 = 4.875$, yielding a rate $R = 1/4.875 \approx 0.205$. 

The privacy of the protocol can be verified carefully by running through arguments as in Appendix \ref{app:ProofofPrivacy_PIRgeneralgraph}. The essential property of the construction that ensures privacy is (a) the variables $X_i,Y_i$ are mutually independent and independent of $\theta$, and (b) since the `shift' is applied to a upstream Bernoulli query variable  when constructing the query for the downstream server corresponding to the desired file, and this variable is independent of all others that appear in the query of the downstream server. 
\end{example}

\section{Discussion}
\label{sec:conclusion}
In this work, we have presented two variable-download schemes, one specialized for the star graphs and another for any graph. Our schemes operate with $L=1$ which is the best-case subpacketization. Our schemes essentially exploit the event of seeing a null-query, which necessitates no response. This technique was previously employed in \cite{tian2019capacity} to obtain a capacity-achieving PIR scheme with optimal message-size, for the case of fully replicated databases. Our work shows the effective of this technique for graph-based replicated storage as well. We believe that, with further improvements, this technique could lead to capacity-achieving schemes with minimal subpacketization, at least for some special classes of graphs. 
% Over general graphs,  we were able to send the null query with probability half for a subset of the servers. And over star graphs, We increase this probability by a lot. 
%%%%%%%%%%%%%%%%%%%%%%%
% \IEEEtriggeratref{11}
\bibliographystyle{IEEEtran}
\bibliography{graphs_PIR_Fullversion_ArXiv_single}

@ARTICLE{ravivetal_2018_TIT,
  author={Raviv, Netanel and Tamo, Itzhak and Yaakobi, Eitan},
  journal={IEEE Transactions on Information Theory}, 
  title={Private Information Retrieval in Graph-Based Replication Systems}, 
  year={2020},
  volume={66},
  number={6},
  pages={3590-3602},
  doi={10.1109/TIT.2019.2955053}}

@ARTICLE{attia_kumar_tandon_storageconstrained_TIT,
  author={Attia, Mohamed Adel and Kumar, Deepak and Tandon, Ravi},
  journal={IEEE Transactions on Information Theory}, 
  title={The Capacity of Private Information Retrieval From Uncoded Storage Constrained Databases}, 
  year={2020},
  volume={66},
  number={11},
  pages={6617-6634},
  doi={10.1109/TIT.2020.3023016}}

@ARTICLE{tPIRIEEE,
  author={Freij-Hollanti, Ragnar and Gnilke, Oliver W. and Hollanti, Camilla and Horlemann-Trautmann, Anna-Lena and Karpuk, David and Kubjas, Ivo},
  journal={IEEE Transactions on Information Theory}, 
  title={$t$-private information retrieval schemes using transitive codes}, 
  year={2019},
  volume={65},
  number={4},
  pages={2107-2118},
  doi={10.1109/TIT.2018.2871050}}

@ARTICLE{PIR_linearcodes_arbit,
  author={Kumar, Siddhartha and Lin, Hsuan-Yin and Rosnes, Eirik and Graell i Amat, Alexandre},
  journal={IEEE Transactions on Information Theory}, 
  title={Achieving Maximum Distance Separable Private Information Retrieval Capacity With Linear Codes}, 
  year={2019},
  volume={65},
  number={7},
  pages={4243-4273},
  doi={10.1109/TIT.2019.2900313}}

@article{FriejHollantiSIAM,
author = {Freij-Hollanti, Ragnar and Gnilke, Oliver W. and Hollanti, Camilla and Karpuk, David A.},
title = {Private Information Retrieval from Coded Databases with Colluding Servers},
journal = {SIAM Journal on Applied Algebra and Geometry},
volume = {1},
number = {1},
pages = {647-664},
year = {2017},
doi = {10.1137/16M1102562},
URL = {https://doi.org/10.1137/16M1102562},
eprint = {https://doi.org/10.1137/16M1102562}}

@INPROCEEDINGS{2020_IWCIT_OptimalMsgSizePIRnonrepl_keramaatietal,
  author={Keramaati, S. Niloofar and Salehkalaibar, Sadaf},
  booktitle={2020 Iran Workshop on Communication and Information Theory (IWCIT)}, 
  title={Private Information Retrieval from Non-Replicated Databases with Optimal Message Size}, 
  year={2020},
  volume={},
  number={},
  pages={1-6},
  doi={10.1109/IWCIT50667.2020.9163525}}

@ARTICLE{Kong_TIT_2025newcapacityboundspir,
  author={Kong, Xiangliang and Meel, Shreya and Maranzatto, Thomas Jacob and Tamo, Itzhak and Ulukus, Sennur},
  journal={IEEE Transactions on Information Theory}, 
  title={New Capacity Bounds for PIR on Graph and Multigraph-Based Replicated Storage}, 
  year={2026},
  volume={72},
  number={1},
  pages={691-709},
  doi={10.1109/TIT.2025.3631441}}

@INPROCEEDINGS{Yuhangetal_Capof4StarGraph_ISIT_2023,
  author={Yao, Yuhang and Jafar, Syed A.},
  booktitle={2023 IEEE International Symposium on Information Theory (ISIT)}, 
  title={The Capacity of 4-Star-Graph PIR}, 
  year={2023},
  volume={},
  number={},
  pages={1603-1608},
  doi={10.1109/ISIT54713.2023.10206729}}

@misc{ge2025privateinformationretrievalgraphs,
      title={Private Information Retrieval over Graphs}, 
      author={Gennian Ge and Hao Wang and Zixiang Xu and Yijun Zhang},
      year={2025},
      eprint={2509.26512},
      archivePrefix={arXiv},
      primaryClass={cs.IT},
      url={https://arxiv.org/abs/2509.26512}, 
}

@misc{meel2025privateinformationretrievalmultigraphbased,
      title={Private Information Retrieval on Multigraph-Based Replicated Storage}, 
      author={Shreya Meel and Xiangliang Kong and Thomas Jacob Maranzatto and Itzhak Tamo and Sennur Ulukus},
      year={2025},
      eprint={2501.17845},
      archivePrefix={arXiv},
      primaryClass={cs.IT},
      url={https://arxiv.org/abs/2501.17845}, 
}

@ARTICLE{Sadehetal_TIT_2023_BoundsonPIRgraphs,
  author={Sadeh, Bar and Gu, Yujie and Tamo, Itzhak},
  journal={IEEE Transactions on Information Forensics and Security}, 
  title={Bounds on the Capacity of Private Information Retrieval Over Graphs},
  year={2023},
  volume={18},
  number={},
  pages={261-273},
  doi={10.1109/TIFS.2022.3220034},
}

@ARTICLE{Tian2020_storageCost,
  author={Tian, Chao},
  journal={IEEE Transactions on Information Theory}, 
  title={On the Storage Cost of Private Information Retrieval}, 
  year={2020},
  volume={66},
  number={12},
  pages={7539-7549},
  doi={10.1109/TIT.2020.3015818}}

@ARTICLE{tian2019capacity,
  author={Tian, Chao and Sun, Hua and Chen, Jun},
  journal={IEEE Transactions on Information Theory}, 
  title={Capacity-Achieving Private Information Retrieval Codes With Optimal Message Size and Upload Cost}, 
  year={2019},
  volume={65},
  number={11},
  pages={7613-7627},
  doi={10.1109/TIT.2019.2918207}}

@ARTICLE{sun2017capacityb,
  author={Sun, Hua and Jafar, Syed Ali},
  journal={IEEE Transactions on Information Theory}, 
  title={The Capacity of Robust Private Information Retrieval With Colluding Databases}, 
  year={2018},
  volume={64},
  number={4},
  pages={2361-2370},
  doi={10.1109/TIT.2017.2777490}}

@ARTICLE{sun2017capacity,
  author={Sun, Hua and Jafar, Syed Ali},
  journal={IEEE Transactions on Information Theory}, 
  title={The Capacity of Private Information Retrieval}, 
  year={2017},
  volume={63},
  number={7},
  pages={4075-4088},
  doi={10.1109/TIT.2017.2689028}}

@ARTICLE{banawan2018capacity,
  author={Banawan, Karim and Ulukus, Sennur},
  journal={IEEE Transactions on Information Theory}, 
  title={The Capacity of Private Information Retrieval From Coded Databases}, 
  year={2018},
  volume={64},
  number={3},
  pages={1945-1956},
  doi={10.1109/TIT.2018.2791994}}

@article{chor1998_PIR,
author = {Chor, Benny and Kushilevitz, Eyal and Goldreich, Oded and Sudan, Madhu},
title = {Private information retrieval},
year = {1998},
issue_date = {Nov. 1998},
publisher = {Association for Computing Machinery},
address = {New York, NY, USA},
volume = {45},
number = {6},
issn = {0004-5411},
url = {https://doi.org/10.1145/293347.293350},
doi = {10.1145/293347.293350},
journal = {J. ACM},
month = nov,
pages = {965–981},
numpages = {17}
}
\newpage
\appendices
\section{Privacy of PIR Protocol in Section \ref{sec:generalgraph1} for General Graphs}
\label{app:ProofofPrivacy_PIRgeneralgraph}
To establish privacy, we must demonstrate that for any server $n \in V$, the  query  $Q_n$ is independent of the client's desired file index, $\theta$. 

To proceed, we establish a key property of individual query bits.
%%%%
\begin{claim}
\label{Claim:fjs_are_bernoul}
For any server $m \in V$ and any $j \in [d_m]$, the query bit $f_j^m$ is uniformly distributed on $\Fb_2$, for any given $\theta$. 
\end{claim}
    We now prove this claim (by induction on set index $t=1,\dots,\kappa$). The base case $t=1$ is obvious, as for any $m\in I_1$, $Q_m$ is uniform on $\{\overline{0},\overline{1}\}$. Now, assume the claim holds for all servers in $I_1, \dots, I_{t-1}$. Let $m \in I_t$. If $f_j^m$ is part of the downstream subvector for server $m$, it is uniform on $\Fb_2$ by construction. If it is part of the upstream subvector, then $f_j^m = f_b^p \oplus \mathbb{I}(W_{m,p}=W_\theta)$, for some server $p$ in a preceding set. Here, $\mathbb{I}(W_{m,p}=W_\theta)$ is indicator random variable, distributed on $\{0,1\}=\Fb_2$. By the claim's inductive hypothesis for $p$, $f_b^p$ is a uniform bit. Thus, $f_j^m$ is uniform. Thus, the claim holds.

We now proceed to show that the  query  $Q_n$ is independent of $\theta$ for all $n\in I_1\cup\dots\cup I_{\kappa}$, by  induction on the index $s$ of the independent sets, from $s=1$ to $\kappa$.

\textit{Base Case ($s=1$): } Consider any server $n \in I_1$. According to the protocol, its query vector $Q_n$ is chosen to be the all-zeros vector ($\overline{0}$) or the all-ones vector ($\overline{1}$), each with probability $1/2$. Thus, the $Q_n$ is clearly independent of $\theta$.  

\textit{Inductive Hypothesis:} Now, we assume that for some integer $s \in [2, \kappa]$, the privacy property holds for all servers in the preceding independent sets. That is, for any server $m \in I_1 \cup I_2 \cup \dots \cup I_{s-1}$, we assume $Q_m$ is independent of $\theta$.

\textit{Inductive Step: Proving for $I_s$:} Consider an arbitrary server $n \in I_s$. We need to show that the distribution of its query $Q_n$ is independent of $W_\theta$. The query $Q_n$ is constructed by concatenating the downstream and upstream subvectors: 
\begin{enumerate}
    \item  $Q_n^{\downarrow} = (f_1^n, \dots, f_{d_n^{\downarrow}}^n)$: For edges connecting to servers in $I_{s+1} \cup \dots \cup I_{\kappa}$.
    \item  $Q_n^{\uparrow} = (f_{d_n^{\downarrow}+1}^n, \dots, f_{d_n}^n)$: For edges connecting to servers in $I_1 \cup \dots \cup I_{s-1}$.
\end{enumerate}
%%%%

Now we analyze the subvectors of $Q_n$.

\textit{1. Distribution of $Q_n^{\downarrow}$:}
The subvector $Q_n^{\downarrow}$ is chosen to be $\overline{0}$ or $\overline{1}$ with probability $1/2$. We may visualize that this choice is decided by a random independent fair-coin flip (denote by $C_n$), performed uniquely at server $n$. The distribution of $Q_n^{\downarrow}$ is explicitly independent of $W_\theta$.

\textit{2. Distribution of $Q_n^{\uparrow}$:}
Each bit $f_j^n$ in this subvector corresponds to an edge $(n,m)$, where $m \in I_t$ for some $t < s$. Let the bit in $Q_m$ for this shared file be $f_a^m$. The protocol defines $f_j^n = f_a^m \oplus \mathbb{I}(W_{n,m} = W_\theta)$.

Now, from server $m$'s perspective, the edge $\{m,n\}$ is a downstream edge to server $n$ is in a \textit{later}  $I_s$. Therefore, the bit $f_a^m$ is an upstream bit, whose value is determined by the independent coin flip $C_m$ at server $m$. Further, for any $j_1,j_2\in\{d_n^\uparrow+1,\dots,d_n\}$, the bits $f_{j_1}^n,f_{j_2}^n$ can be written as $f_{j_1}^n = f_{a_1}^{m_1} \oplus \mathbb{I}(W_{n,m_1} = W_\theta)$ and $f_{j_2}^n = f_{a_2}^{m_2} \oplus \mathbb{I}(W_{n,m_2} = W_\theta)$ for \textit{distinct} upstream vertices $m_1,m_2$ in some prior sets $I_{t_1},I_{t_2}$ respectively. The values of $f_{a_1}^{m_2}$ and $f_{a_2}^{m_2}$ are therefore determined by independent coin flips $C_{m_1}$ and $C_{m_2}$ respectively. Thus, each of the $d_n - d_n^{\downarrow}$ upstream bits used to construct $Q_n^{\uparrow}$ are also mutually independent random variables, each uniformly distributed on $\Fb_2$. Clearly, by its structure so described, $Q_n^{\uparrow}$ is also independent of $\theta$, via the induction hypothesis. 

\textit{3. Independence of $Q_n^{\downarrow}$ and $Q_n^{\uparrow}$:} We also observe that the subvectors $Q_n^{\downarrow}$ and $Q_n^{\uparrow}$ themselves must be mutually independent (given $\theta$), due to the independence of coin flips $C_m:m\in (I_1\cup\dots I_{s-1})$ and $C_s$.

Using the points 1,2,3 above together, we have proved that $P_{Q_n|\theta}=P_{Q_n^\downarrow|\theta}P_{Q_n^\uparrow|\theta}=P_{Q_n^\downarrow}P_{Q_n^\uparrow}$. Thus $Q_n$ is independent of $\theta$. 

\end{document}